\documentclass[11pt,a4paper]{article}

\usepackage{amsmath,amssymb,amsfonts}
\usepackage{amsthm}
\usepackage{mathrsfs}

\usepackage{graphicx}
\usepackage{xcolor}

\usepackage{multirow}
\usepackage{booktabs}
\usepackage{tabularx}

\usepackage{algorithm}
\usepackage{algorithmicx}
\usepackage{algpseudocode}

\usepackage[title]{appendix}
\usepackage{textcomp}
\usepackage{listings}
\usepackage{url}
\usepackage[colorlinks=true,linkcolor=blue,citecolor=blue,urlcolor=blue]{hyperref}

\usepackage[margin=1in]{geometry}

\usepackage{setspace}
\usepackage{authblk}

\begin{document}

\title{How Recommendation Algorithms Shape Social Networks: An Adaptive Voter Model Approach}

\author[1]{Fabian Veider}
\author[1]{Georg Jäger}
\author[2]{Bao Quoc Tang}

\affil[1]{Department of Environmental Systems Sciences, University of Graz, Merangasse 18, 8010 Graz, Austria}
\affil[2]{Department of Mathematics and Scientific Computing, University of Graz, Heinrichstrasse 36, 8010 Graz, Austria}

\date{\today}

\maketitle

\begin{abstract}
The rise of social media and recommendation algorithms has sparked concerns about their role in fostering opinion polarization and echo chambers. We study these phenomena using an adaptive voter model to compare two connection mechanisms: "free" global rewiring, where individuals connect with anyone sharing their opinion, and "friend-of-a-friend" local rewiring, which mimics algorithmic link recommendations on platforms like Facebook or LinkedIn. Simulations across different network topologies reveal that local rewiring increases final-state polarization of the system and fragments social networks into many disconnected components. The usual phase transition into two disconnected components turns into a fragmentation of smaller components, leading to an increase in echo chambers as well as many isolated nodes. This effect is most pronounced in clustered networks with high homophily in rewiring, illustrating how recommendation algorithms can intensify social fragmentation by changing the very structure of the network.
\end{abstract}


\medskip

\noindent\textbf{Corresponding author:} Fabian Veider (\texttt{fabian.veider@uni-graz.at})
	
	\section{Introduction}
	The digitization of social interactions through platforms like Facebook, LinkedIn and others has raised concerns about their role in fostering societal divide \cite{prior2013media,lelkes2017hostile}, given the general rise of social media usage \cite{perrin2015social, auxier2021social}. While empirical evidence shows correlations between social media use and polarization \cite{tornberg2021modeling, cinelli2021echo}, the causal mechanisms remain debated. A rising part of research investigates how algorithms can impact the information flow on social media \cite{brinkmann2023machine, wagner2021measuring}, asking the question if and how algorithms can shape opinions. While some case studies indicate that content algorithms may not show increased affective polarization \cite{feezell2021exploring}, a substantial body of research suggests that recommendation algorithms, either via content filtering \cite{perra2019modelling} or link formation \cite{santos2021link, cinus2022effect}, can play a substantial role in creating "filter bubbles" \cite{pariser2011filter} that limit exposure to diverse viewpoints, potentially leading to echo chambers \cite{nielsen2020democratic}, networks that feature both opinion polarization and network segregation. Understanding how they emerge within social media and how they interplay with polarization has recently been studied \cite{cota2019quantifying, diaz2022echo, vendeville2025modeling}, looking into the effects of (mis-)information flow in networks with echo chambers \cite{tornberg2018echo}. Empirical research on echo chambers and polarization often relies on large-scale data analysis, social media experiments, surveys, and increasingly on Natural Language Processing (NLP) and Large Language Models (LLMs) to analyze discourse or infer attitudes. While LLMs offer new possibilities for classification and text analysis, their use as behavioral simulators remains contested \cite{huang2024social,larooij2025validation}. Because such generative models embed opaque training biases and do not reliably reproduce human decision-making processes, they are poorly suited for isolating causal mechanisms of opinion change. These challenges motivate complementary approaches based on mechanistic, interpretable models.
    \newline\newline     Given these empirical constraints, we use a computational modeling approach grounded in opinion dynamics \cite{castellano2009statistical, starnini2025opinion}. This field draws heavily from agent-based simulations to understand how collective phenomena, such as polarization and echo chambers, can emerge from individual interactions. These agent-based models (ABMs) have been instrumental in modeling a wide range of social dynamics, from cultural dissemination \cite{axelrod1997dissemination} to residential segregation \cite{schelling1971dynamic}. Generally, agents are represented as nodes in a network, endowed with opinions and connections representing social ties. The system then evolves through two key processes: agents update their opinions based on other nodes, often their direct neighbors, and they may alter their network connections over time. These dynamics are typically governed by sociological principles like homophily, which is the tendency to be surrounded by similar people \cite{mcpherson2001birds, lee2019homophily}, and social influence, the process of becoming more similar to one's neighborhood \cite{centola2007complex, turner1991social}. Crucial for digital social networks, one has to also consider algorithm effects, formalized to mimic both content-selection or mechanisms of link rewiring \cite{cinus2022effect, santos2021link}. A broad class of ABMs within this framework feature one-dimensional continuous opinions \cite{chitra2020analyzing, baumann2020modeling,valensise2023drivers,santos2021link, ninomiya2025mitigating} that represent individuals' stance towards a political topic (partisanship, policy approval or other socio-political traits). While these continuous opinion models are highly effective for capturing the spectrum of many beliefs, a substantial number of opinion measures are represented discretely. Electoral choices, referendum outcomes, and major policy decisions often present binary options that individuals need to decide on, and inherently nuanced opinions are frequently measured via discrete surveys using a Likert scales \cite{berinsky2017measuring} (with some exceptions \cite{reips2008interval}).\newline\newline 
    In such contexts, discrete opinion models like the voter model (VM) still provide a valuable framework for understanding polarization dynamics \cite{liggett2013stochastic}, which is implemented via a simple rule of social influence: an agent adopts the opinion of a randomly selected neighbor \cite{castellano2003incomplete, suchecki2004conservation}. This mechanism provides a foundational model for how consensus can emerge from local interactions \cite{holley1975ergodic}. The adaptive voter model (AVM) extends on the VM by incorporating link rewiring \cite{mcpherson2001birds}, allowing the network structure to co-evolve with opinions \cite{vazquez2008generic, holme2006nonequilibrium}. This flexibility has led to AVMs being applied to case studies of polarization through news sources and algorithmic personalization \cite{bhat2019opinion, perra2019modelling}, as well as other contexts such as swarming behavior \cite{zschaler2012adaptive} and epidemics \cite{gross2008adaptive}. A closely related strand of research focuses on triadic closure \cite{klimek2016dynamical, mosleh2025tendencies}, where agents form links to friends-of-friends independent of opinion. In contrast to this opinion-agnostic mechanism, our model investigates a simple local rewiring mechanism inspired by social media algorithms. To analyse this, we use an AVM to expand on research of polarization and echo chambers, allowing us to isolate the influence of platform design from unknown factors. Our approach offers a mechanistic understanding of polarization in digital social networks through a simulated society \cite{gilbert2018simulating} that can be used and extended to test alternatives to current social media landscapes \cite{larooij2025can}.\newline\newline
    Our findings reveal that overall polarization increases for algorithmic rewiring and fundamentally alters network connectedness. It produces numerous small echo chambers and a smooth fragmentation process, in contrast to the sharp phase transition observed under standard rewiring. These effects appear across all examined network topologies, underscoring the sensitivity of echo-chamber formation and connectivity to platform design.
    \newline\newline
	In Section \ref{Methodology} we outline the rules and setup for our analysis. Section \ref{Results} showcases all results for different network topologies, focusing on the temporal evolution of polarization measures as well as echo chamber distributions and measures of criticality. We summarize our results and conclude with an outlook for future research in Section \ref{Conclusions}. Additional simulation results emphasizing the robustness of our results can be found in the appendix.
	
	\section{Model}\label{Methodology}
	We implement an adaptive voter model (AVM) where each agent $i=1,...,N$ holds a binary opinion $x_i \in \{0,1\}$, representing a voting choice, topic preference or another cultural trait. At $t=0$, opinions are assigned at random such that exactly half of the agents hold opinion 0 and half hold opinion 1, with each realization being a different random configuration. The model dynamics proceed in discrete timesteps, with each step consisting of the following process:
	
	\newpage
	
	\begin{enumerate}
		\item A random agent $i$ is selected uniformly from the population of $N$ agents
		\item With probability $p_s,~s\in\{g,l\}$, the agent rewires one of its existing links to a new target agent with matching opinion. We distinguish two types of rewiring scenarios:
        \begin{itemize}
    		\item \textbf{Global homophilic rewiring $p_g$}: A link is reconnected to any randomly selected agent in the network that shares the same opinion.
    		\item \textbf{Local homophilic rewiring $p_{l}$}: A link is reconnected to a neighbor of a neighbor that shares the same opinion, mimicking the "friends-of-friends" recommendation algorithms used in social platforms. No rewiring occurs if no like-minded node can be found during that timestep.
    	\end{itemize}
		\item With probability $1-p_s$, the agent imitates a randomly selected neighbor $j$ and adopts their opinion $x_j$. 
	\end{enumerate}
	\noindent One obtains the direct voter model for the special case $p_s=0$. The rewiring probability $p_s$ is varied from $p_s=0$ to $p_s=1$ (pure rewiring) to capture different strengths of preferential connections. Additionally, we investigate the effect of reducing homophily both for global and local rewiring in order to understand if and how it impacts our definition of polarization and echo chamber formation, allowing us disentangle the effects of homophily and locality in link adjustment. Formally, we do this by adjusting step 2 with another homophily probability $p_h$:
    \begin{enumerate}
		\item[2a.] With probability $p_sp_h,~\{g,l\}$, the agent rewires one of its existing links to a new target with matching opinion
        \item[2b.] With probability $p_s(1-p_h)$, the new  target is chosen from agents with mismatching opinions. 
	\end{enumerate}
	\noindent For $p_h = 1$ this simplifies to the process above. We test the cases $p_h \in [0.9, 0.7, 0.5]$, ranging from high homophily ($p_h = 0.9$) to no homophily ($p_h = 0.5$), bearing resemblance to prior studies on homo- and heterophilic interactions in case of global rewiring \cite{kimura2008coevolutionary}. 
	
	\begin{figure}[H]
		\centering
		\includegraphics[width=0.95\textwidth]{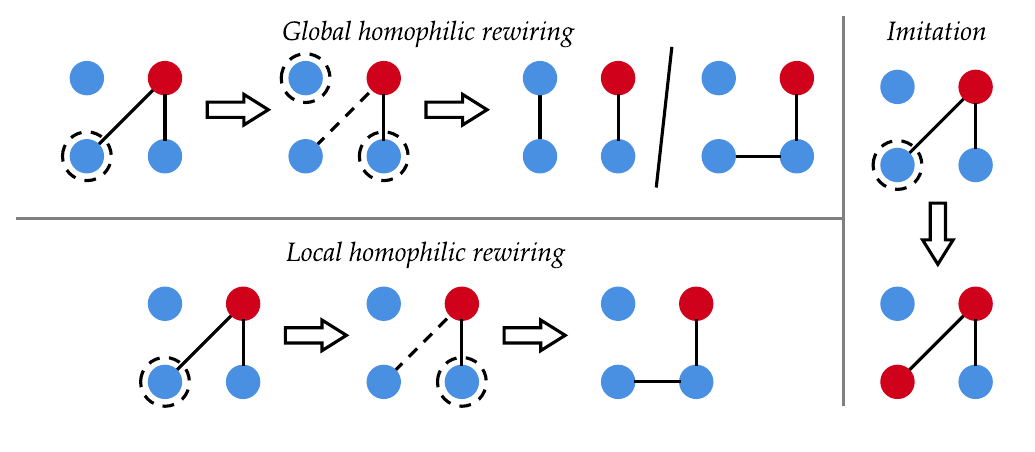}
		\caption{Examples of the different rewiring and opinion update dynamics in a minimal setup. Each color represents a different opinion (for example red = 1, blue = 0). Global rewiring here leads to two equally likely outcomes if the bottom right node is selected, for the other two processes only one update choice exists.}
		\label{fig:dynamics}
	\end{figure}
\begin{figure}[h!]
			\centering
			\includegraphics[width=0.85\linewidth]{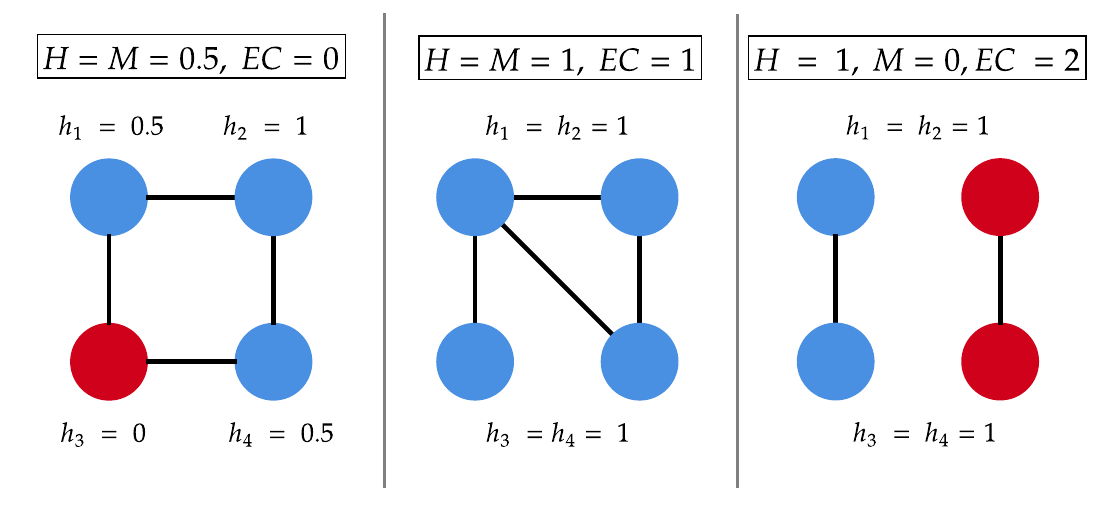}
			\caption{Homophily measure $H$ with individual homophily measures $h_{1-4}$ magnetization $|M|$ and the number of echo chambers $EC$ for three simple graphs with arbitrary opinion assignment (for example, red = 1 and blue = 0). Consensus within the graph or disconnected subgraphs both corresponds to maximal homophily with $H = 1$, as all neighbors share the same opinion in both cases. Magnetization allows us to distinguish between those cases.}
			\label{fig:polarization_overview}
		\end{figure}

	\noindent We decided to use the standard node-update scheme, where each agent interacts with the same probability. Given that the link-update scheme tends to overrepresents highly-connected users and the inverse-update scheme links more closely to actively persuasive agents, this choice seems to be the best fit for our study of digital social media landscapes. This update scheme is then paired with our proposed rewiring dynamics to mimic the active curation of social neighborhoods on friendship-based platforms like Facebook, LinkedIn and Snapchat, with indicators of echo chambers being strongest on Facebook \cite{cinelli2021echo, de2021no}. In this interpretation, global rewiring reflects a user’s unrestricted choice to disconnect and reconnect to any like-minded individual in the network, independent of algorithmic guidance, whereas local rewiring captures recommendation-driven link formation through “friend-of-friend” or neighborhood-based suggestions. Together, these mechanisms approximate how users and platform algorithms jointly reshape online social environments and thus modulate exposure to diverse opinions.
	\newline\newline
	The simulation continues until equilibrium is reached, defined here as the state where no further opinion changes occur. In the special case of $p_{g,l} = 1$, we run simulations until each node has neighbors of their opinion or cannot find new neighbors. For each parameter set, we perform 500 independent realizations to account for stochasticity, and we consider averages and standard deviations for all measures investigated. All simulation results are shown for graphs of size $N=100$, with tests for bigger networks ($N=200,~400$) qualitatively being the same. Topological effects on the dynamics are addressed by applying the model above to three different network types: Erdős-Rényi (ER), Watts-Strogatz (WS) and Barabási-Albert (BA). While the ER network provides a null model for random interactions, real-world social networks exhibit either small-world clustering like the WS network or scale-free degree distributions as seen in the BA network. The WS network is generated with a rewiring probability $p_{wire} = 0.05$ to preserve high clustering \cite{watts1998collective}, while BA networks grow with $m = k_{avg}/2$ links per new node to obtain the same average degrees to compare with other topologies. Key measures in this study include:
	
	\begin{itemize}
		\item \textbf{Homophily $H$}: Quantified at both individual and network level. For each non-isolated node $i$, we first calculate the individual homophily as 
		\begin{equation}
			h_i(t) = \frac{k_{i,s}(t)}{k_i(t)}
		\end{equation}
		where $k_i$ is the node's degree and $k_{i,s}$ counts neighbors sharing its opinion \cite{interian2018empirical}. For network-wide homophily $H$ we average over individual homophily  
		\begin{equation}
			H(t) = \frac{1}{N_{ni}(t)}\sum_i h_i(t)
		\end{equation}
		
		where $N_{ni}(t)$ counts all non-isolated nodes at time $t$, with $H=0$ correspond to all nodes having neighbors with opposing opinions and $H=1$ to all neighbors having the same opinion. We exclude isolated nodes since homophily is undefined for them ($k_i=0$), and assigning arbitrary values would bias the network-level measure, particularly when isolated nodes are numerous.
        \item \textbf{Echo chambers $EC$}: The number of echo chambers ($EC$) are here defined as disconnected subgraphs containing only one opinion and at least two nodes. This incorporates both the idea of like-minded neighbors and network fragmentation as ingredients for echo chambers. We calculate the number of $EC$ and the (normalized) size distribution $s_i = S_i/N$ of disconnected components $S_i$. 
		\item \textbf{Magnetization $M$}: In analogy to physical models of magnetization, we quantify the global opinion balance of the system by
        \begin{equation}
        	M(t) = \frac{N_1(t) - N_0(t)}{N},
        \end{equation}
        where $N_1(t)$ and $N_0(t)$ denote the number of agents holding opinion $1$ and opinion $0$ at time $t$, respectively. Magnetization takes values in the interval $[-1,1]$, with $M=0$ corresponding to an equal split between the two opinions and $|M|=1$ indicating full consensus. While homophily and echo chamber measures capture structural segregation, magnetization directly reflects opinion dominance at the population level and allows us to distinguish polarized but balanced configurations from outcomes in which one opinion prevails. Extending on prior studies \cite{perra2019modelling}, we call social networks polarized if $|M| < 1$ and $H > 0.5$. Maximal polarization occurs in case $|M| = 0$ and $H = 1$ as this means both opinions are equally present but condensed in groups of like-minded people, see the third example in figure (\ref{fig:polarization_overview}).
		\item \textbf{Inequality}: In order to measure inequality in connectivity we calculate the degree distribution $k_i$.
		\item \textbf{Convergence dynamics}: Characterized by both the (normalized) convergence time $\tau=t_{conv}/N$ and the size of the largest component $s_1$. This allows us to investigate critical phase transitions prominent in simple AVMs \cite{holme2006nonequilibrium, vazquez2008generic}.
		
	\end{itemize}

    \noindent The combination of these measures allows us to gain a thorough understanding of network fragmentation and polarization of our similated social networks.
	
	\section{Results \& Discussion} \label{Results}
	
	\begin{figure}[h!]
		\centering
		\includegraphics[width=0.99\linewidth]{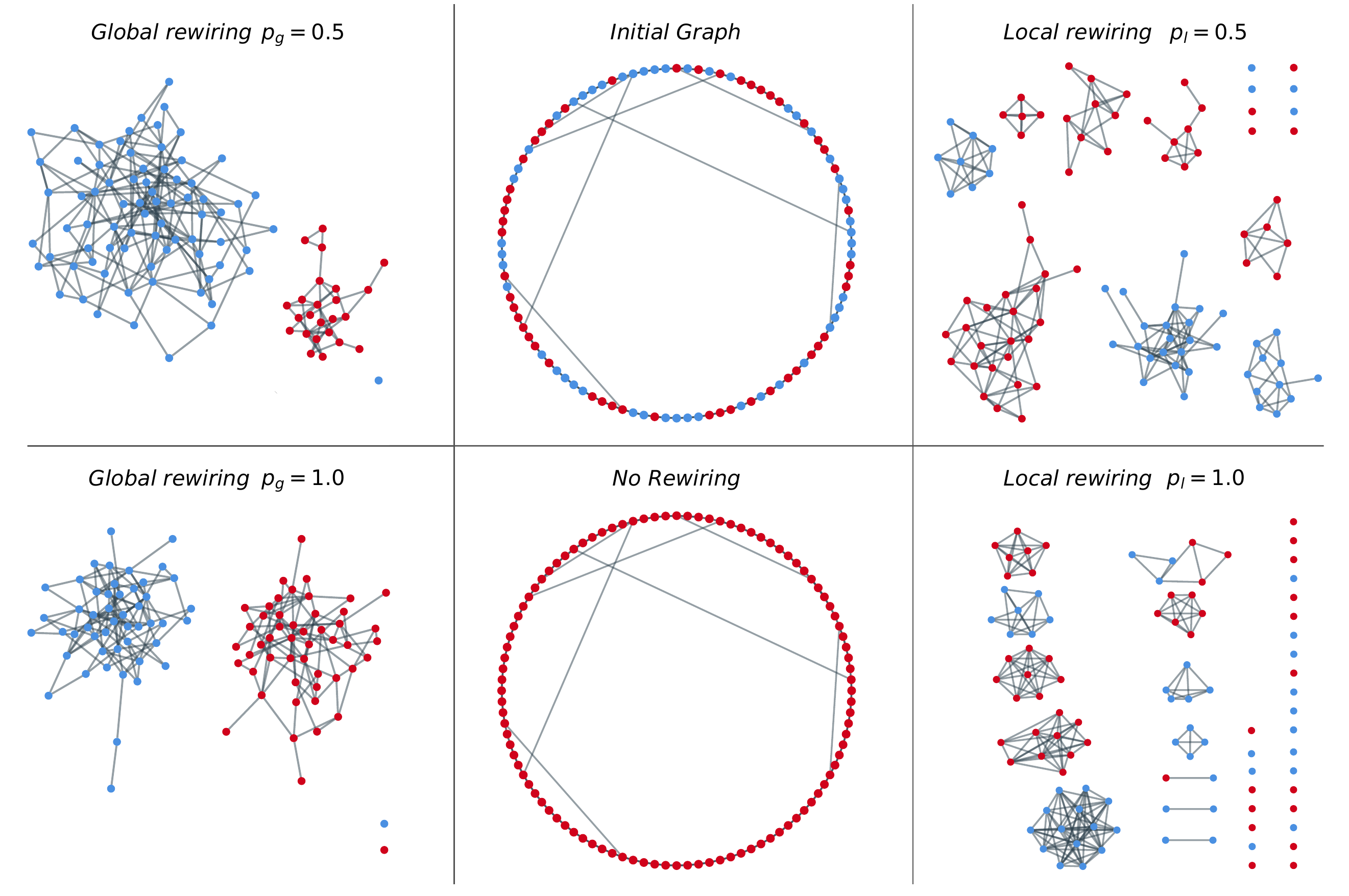}
		\caption{Overview of the two homophilic rewirings for a WS network with $N = 100$. The initial graph (top middle) fragments into multiple smaller components (top left and top right) for local rewiring $p_{l}>0$, compared to two main components (bottom left and bottom right) for global rewiring $p_{g}$. Imitation only (bottom middle) leads to consensus.}
		\label{fig:overview_WS}
	\end{figure}
	
\begin{figure}[H]
		\centering
		\includegraphics[width=0.88\linewidth]{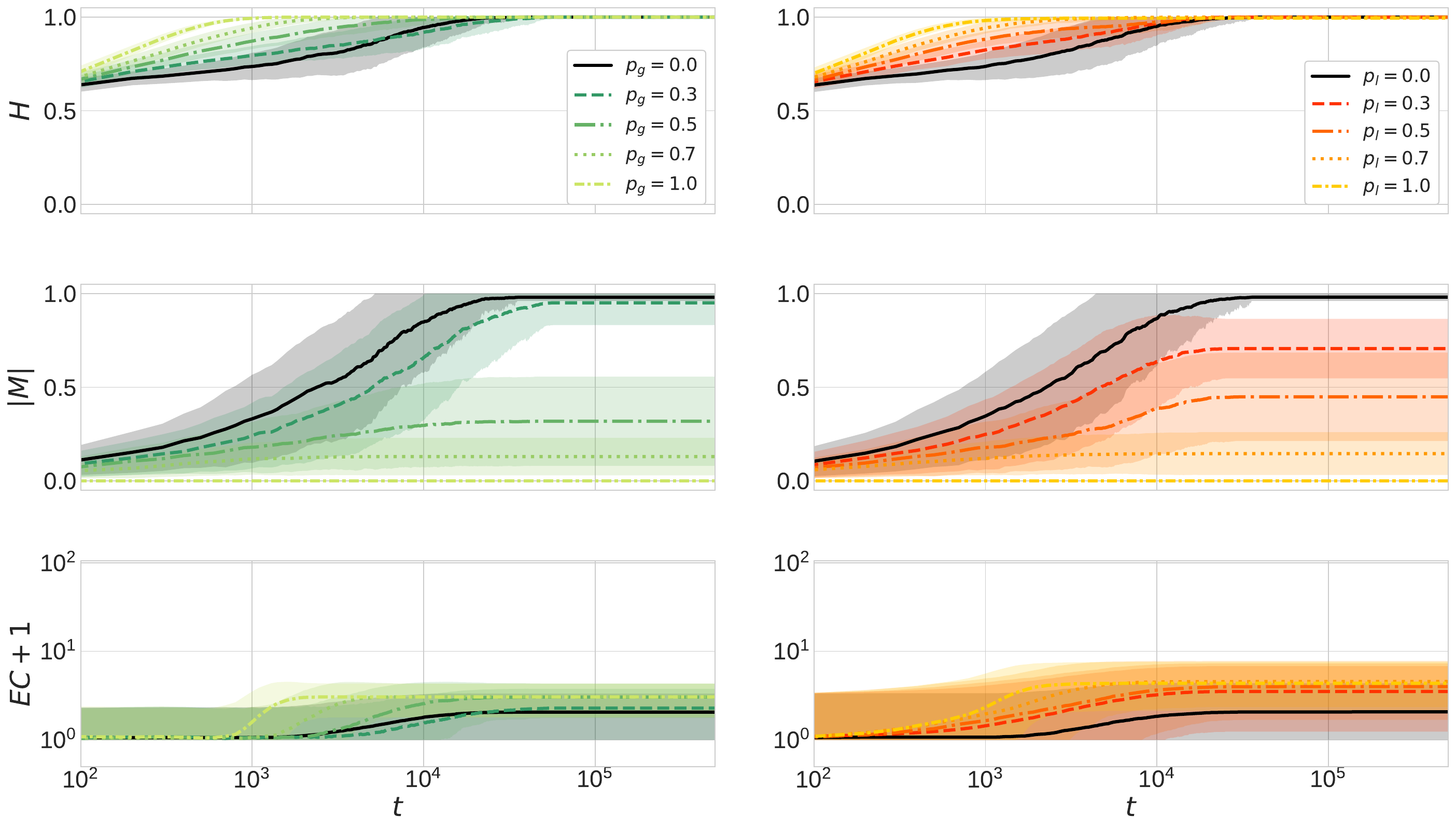}
		
		\includegraphics[width=0.88\linewidth]{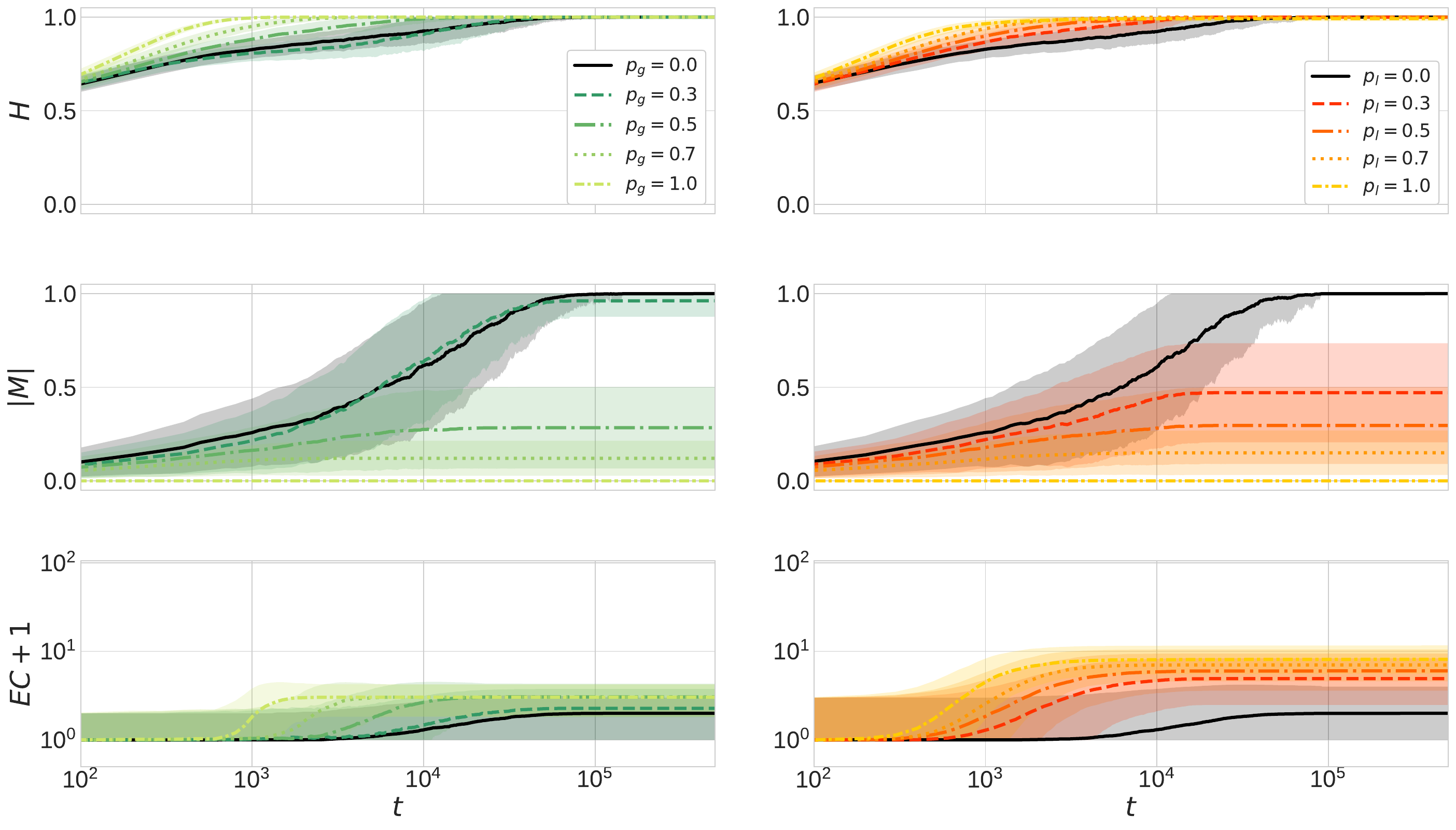}
		\caption{Overall homophily $H$, absolute value of the overall magnetization $|M|$ and the number of echo chambers (shifted by one) $EC+1$ over time $t$ for an ER (top four) and a WS network (bottom four) with $N = 100$ and average degree $k_{avg} = 4$ for different rewiring probabilities. The measures of the BA network (not shown) are essentially the same for the ER network.}
		\label{fig:polarization}
	\end{figure}

	\noindent Figure (\ref{fig:overview_WS}) shows the spectrum of final states for a WS network. For local rewiring $p_l$, a shattered fragmentation of components occurs. In the case of pure rewiring ($p_l = 1$), disconnected components can emerge that contain different opinions, especially for low average degrees (see appendix \ref{appendix:A} for details). The time evolutions of overall homophily $H$, the absolute value of magnetization $|M|$ and the number of echo chambers $EC$ in figure (\ref{fig:polarization}) emphasizes the significant differences between rewiring mechanisms. Local rewiring leads to more echo chambers, especially for the WS network. The system always converges to homophilic neighborhoods as indicated by $H$, but the overall opinion distribution varies between types of rewiring. Generally, local rewiring reduces the final-state magnetization $|M|$, with a more pronounced effect for the WS network. Based on our definition, this corresponds to an increase of polarization with an increase of rewiring that is more pronounced for $p_l$. The equilibration time of all three quantities decreases with increasing $p_{g, l}$, with the ER and BA network sharing extremely similar behavior. The change in convergence time for the BA network based on the skewed selection of high-degree nodes network has been shown in  \cite{barabasi1999emergence} and seems to generalize for local rewiring. Increasing the average degree $k_{avg}$ reduces the number of echo chambers as neighborhood effects decrease for highly connected networks (see appendix \ref{appendix:C}).
	
	\clearpage
	
	\noindent As our choice of a group-based homophily is not unique \cite{musco2021quantify} we tested an alternative quantity for the total homophily $\tilde{H}(t) = \frac{\sum_i k_{i,s}(t)}{\sum_i k_i(t)}$ that sums up all like-minded neigbors and divides by the total number of neighbors. This measure yields qualitatively similar results to the ones in figure (\ref{fig:polarization_overview}). Reducing the rewiring homophily through the parameter $p_h$ creates mostly isolated nodes for local rewiring and overall increased convergence times, see appendix \ref{appendix:D}. This occurs already for strong homophilic rewiring with $p_h = 0.9$, but the overall increase in EC for local rewiring compared to global rewiring still persists, as well as an increased polarization for $p_h < 1$ in case of local rewiring. This signifies a non-trivial interplay between clustering, homophily and local rewiring as drivers for polarization and echo chambers.
	
	\begin{figure}[H]
		\centering
		\includegraphics[width=0.78\linewidth]{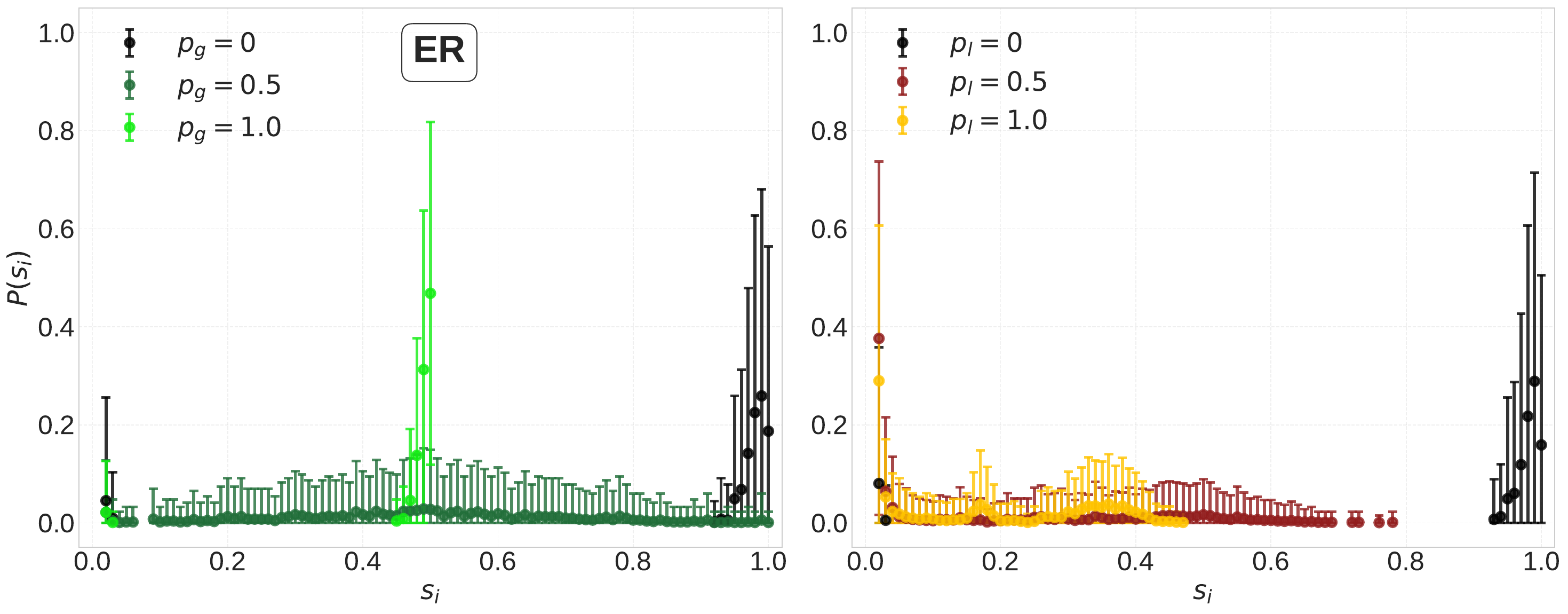}
		\includegraphics[width=0.78\linewidth]{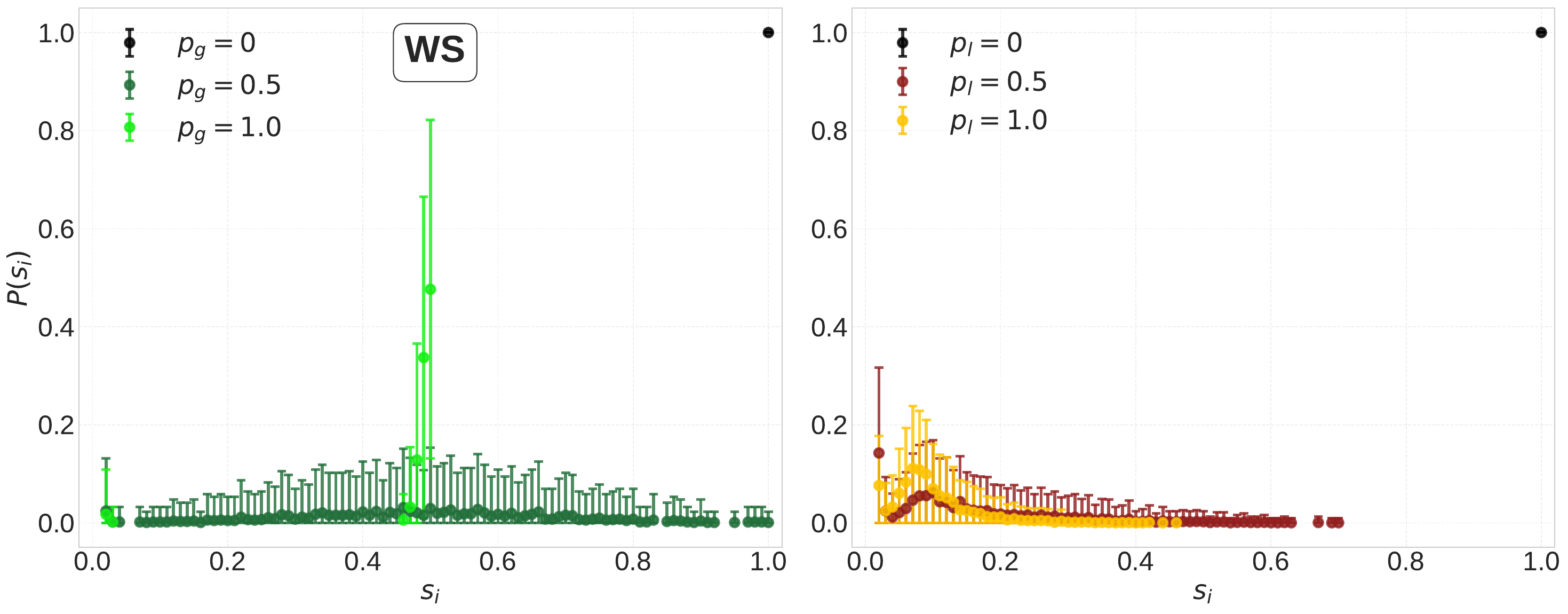}
		\includegraphics[width=0.78\linewidth]{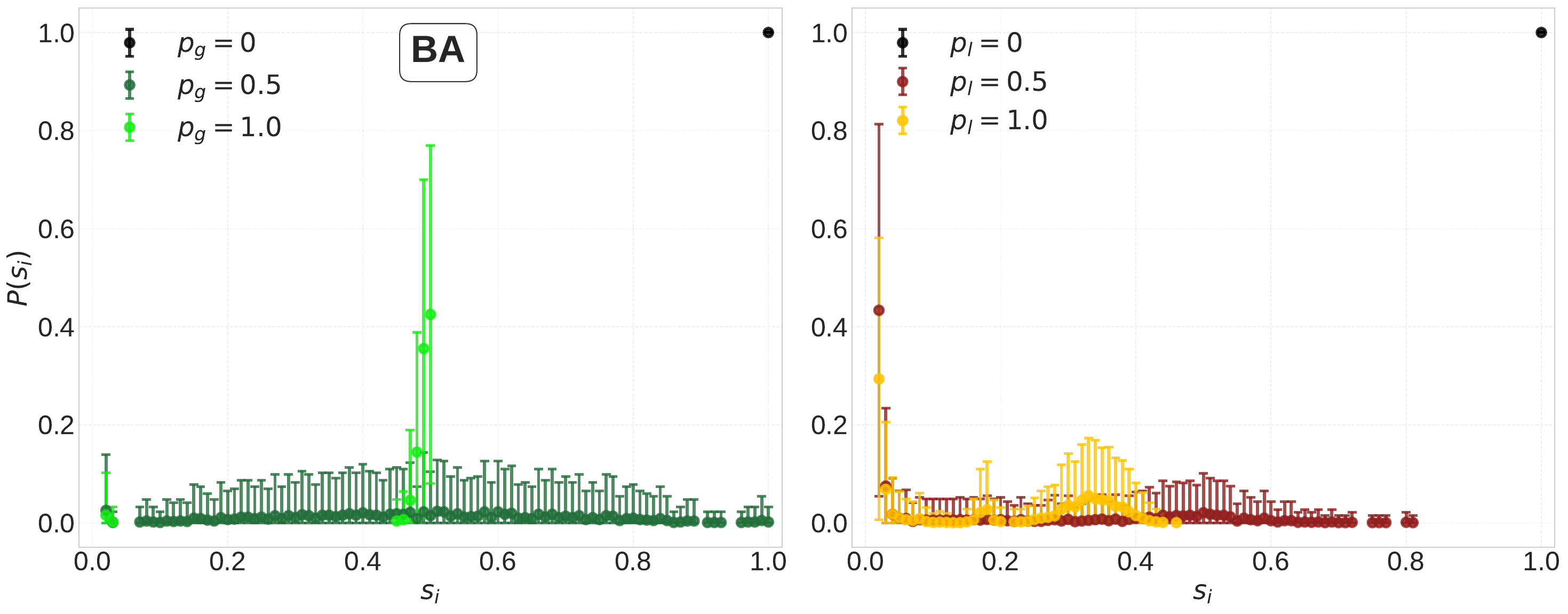}
		\caption{Echo chamber distribution $P(s_i)$ of disconnected subgraphs $s_i$ for an ER (top row), a WS (middle row) and a BA network (bottom row) with $N = 100$ and $k_{avg} = 4$, excluding isolated nodes. In case of no rewiring $p_g = p_l = 0$ only one big cluster exists, indicated by the datapoint in the top right corner in each subfigure.}
		\label{fig:clusters}
	\end{figure}
	
\begin{figure}[h!]
		\centering
		\includegraphics[width=0.85\linewidth]{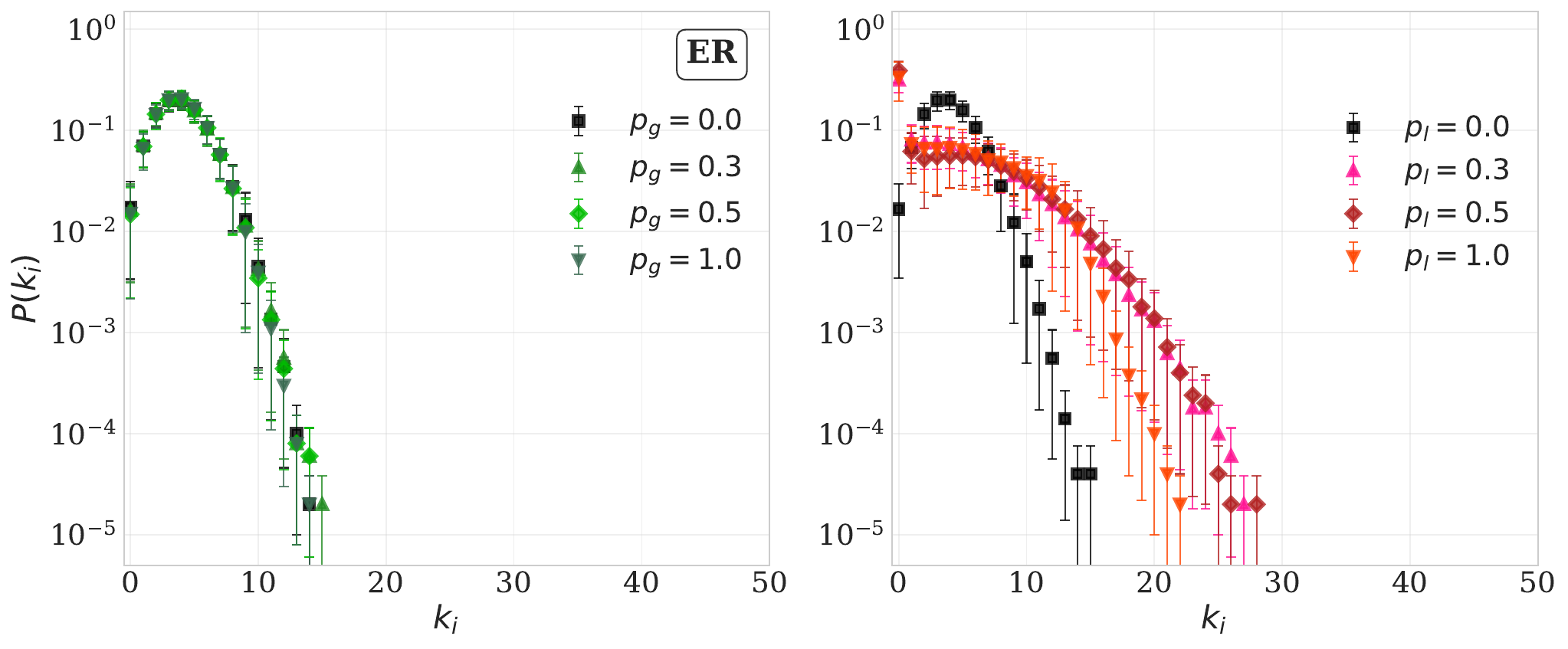}
		\includegraphics[width=0.85\linewidth]{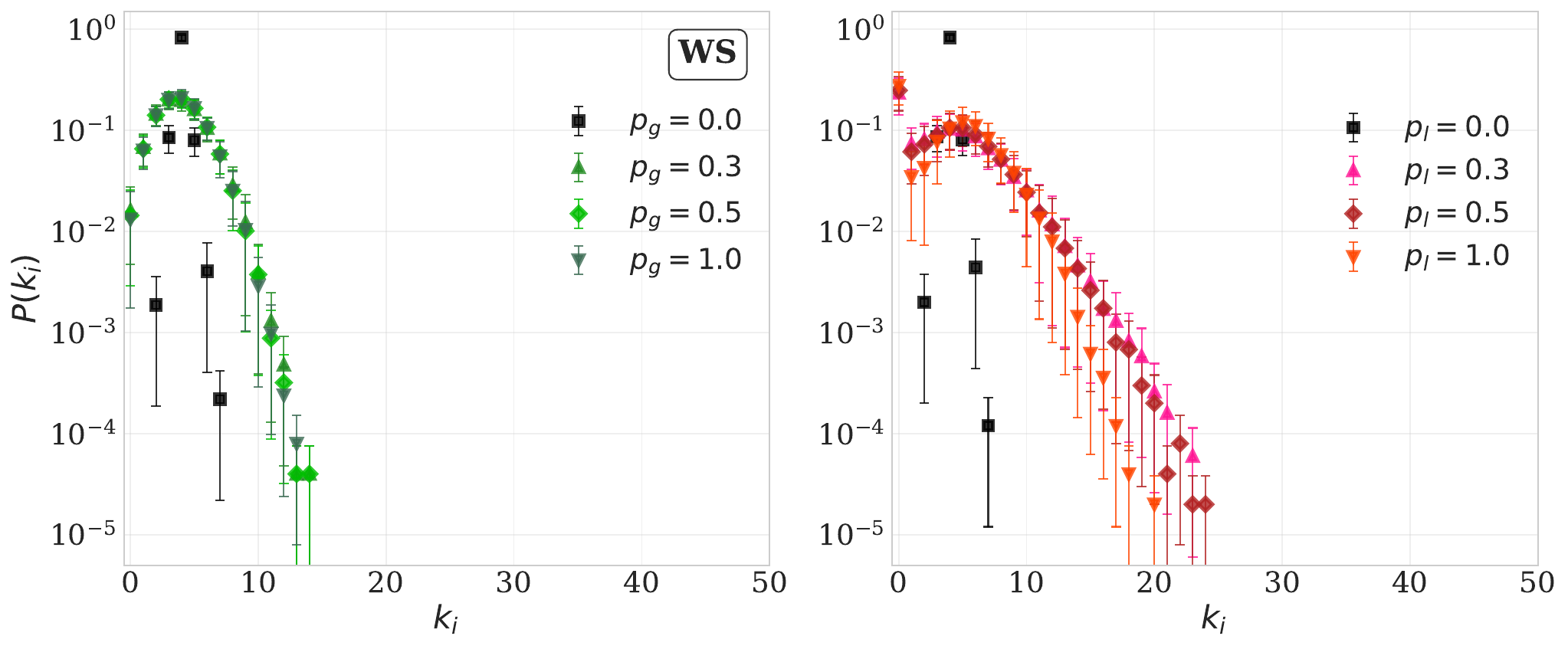}
		\includegraphics[width=0.85\linewidth]{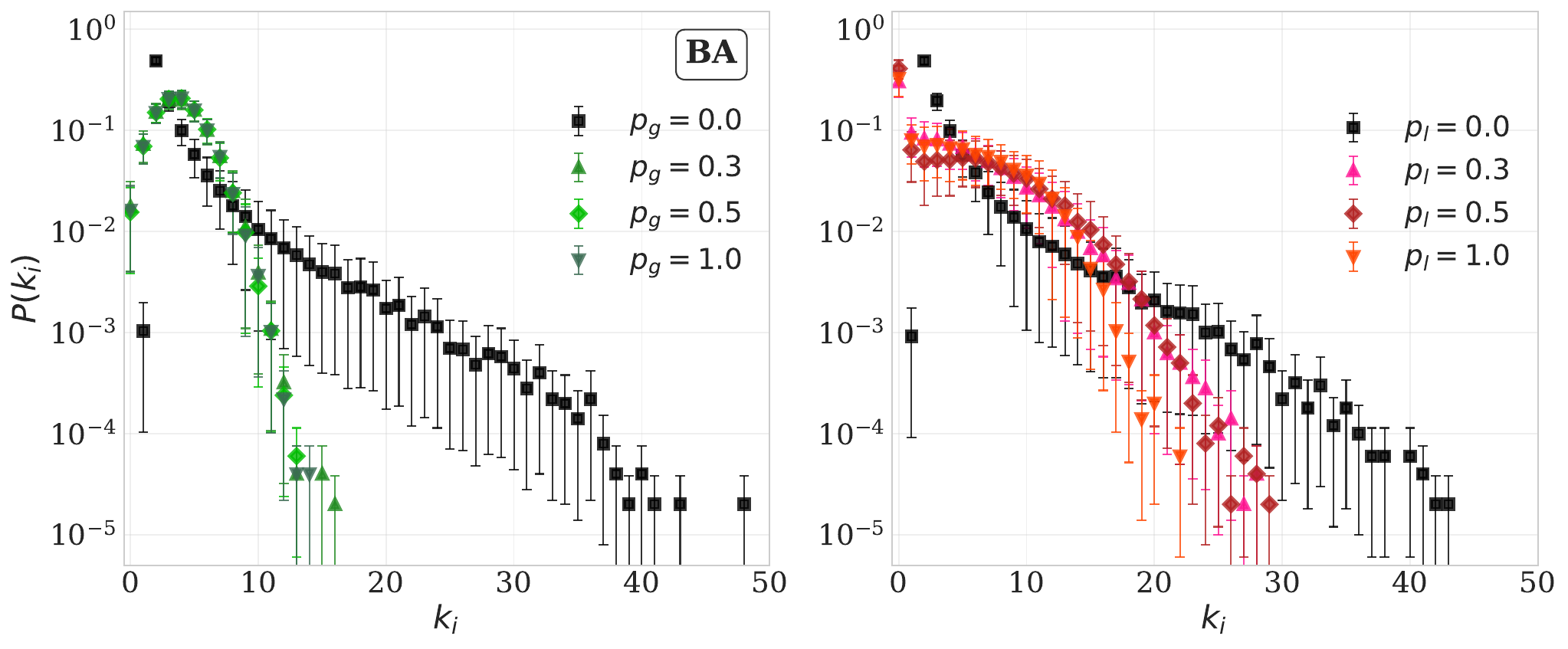}
		\caption{Normalized degree distributions $P(k_i)$ for ER (top row), WS (middle row) and BA (bottom row) network after global rewiring $p_g$ (left) and local rewiring $p_l$ (right) for different rewiring probabilities, with $N=100$ and $k_{avg} = 4$.}
		\label{fig:combineddegreedistributionsk4multiphiwslinear}
	\end{figure}

	\noindent Focusing on the distribution of echo chambers, figure (\ref{fig:clusters}) emphasizes how local rewiring fragments the network into smaller components of various sizes. This fragmentation, normally tied to critical slowing down, has been established for global homophilic rewiring \cite{vazquez2008generic}. For ER networks, isolated nodes in the initialization decreases the size of the biggest component and therefore yields slightly smaller clusters even for $p = 0$. 
	High rewiring probabilities $p_{g,l}$ for all network types yield two clusters of almost equal sizes as we start off with a fifty-fifty distribution of opinions, amplified in case of the WS network. In contrast, local rewiring shifts the size distribution to the left, tending towards many small echo chambers that are furthermore amplified by the clustering of WS networks. The most likely relative size of echo chambers for ER and BA after local rewiring settles around $s_i \approx 0.18$ and $s_i \approx 0.37$, with a significantly smaller size of about $s_i \approx 0.08$ for the WS network. Figure (\ref{fig:combineddegreedistributionsk4multiphiwslinear}) shows that the number of disconnected nodes increases sharply for neighbor rewiring, independent of network type. Additionally, the degree distribution shifts towards the right, with more high-degree nodes emerging for all ranges of $p_l$. BA and ER network yield similar echo chamber distribution for $p_l > 0$, with a slightly smaller minimal size of clusters. This type of shattered fragmentation can also be found in models featuring triadic closure \cite{klimek2016dynamical}, multiplex networks \cite{diakonova2014absorbing} and noise-based update rules \cite{diakonova2015noise}. In contrast, nonlinear variations of adaptive voter model allow for a phase of coexistence instead of this shattered fragmentation \cite{min2017fragmentation, min2023coevolutionary}. Interestingly, the shift towards higher-degree nodes is weakest for the WS network, suggesting a mixed network with the highest number echo chambers but lower inequality in terms of connectedness. This effect is also driven by the fact that, since only small ECs remain, the maximal difference of links cannot be as large compared to bigger ECs.
	
	\newpage
	
\begin{figure}[H]
		\centering
		\includegraphics[width=0.8\textwidth]{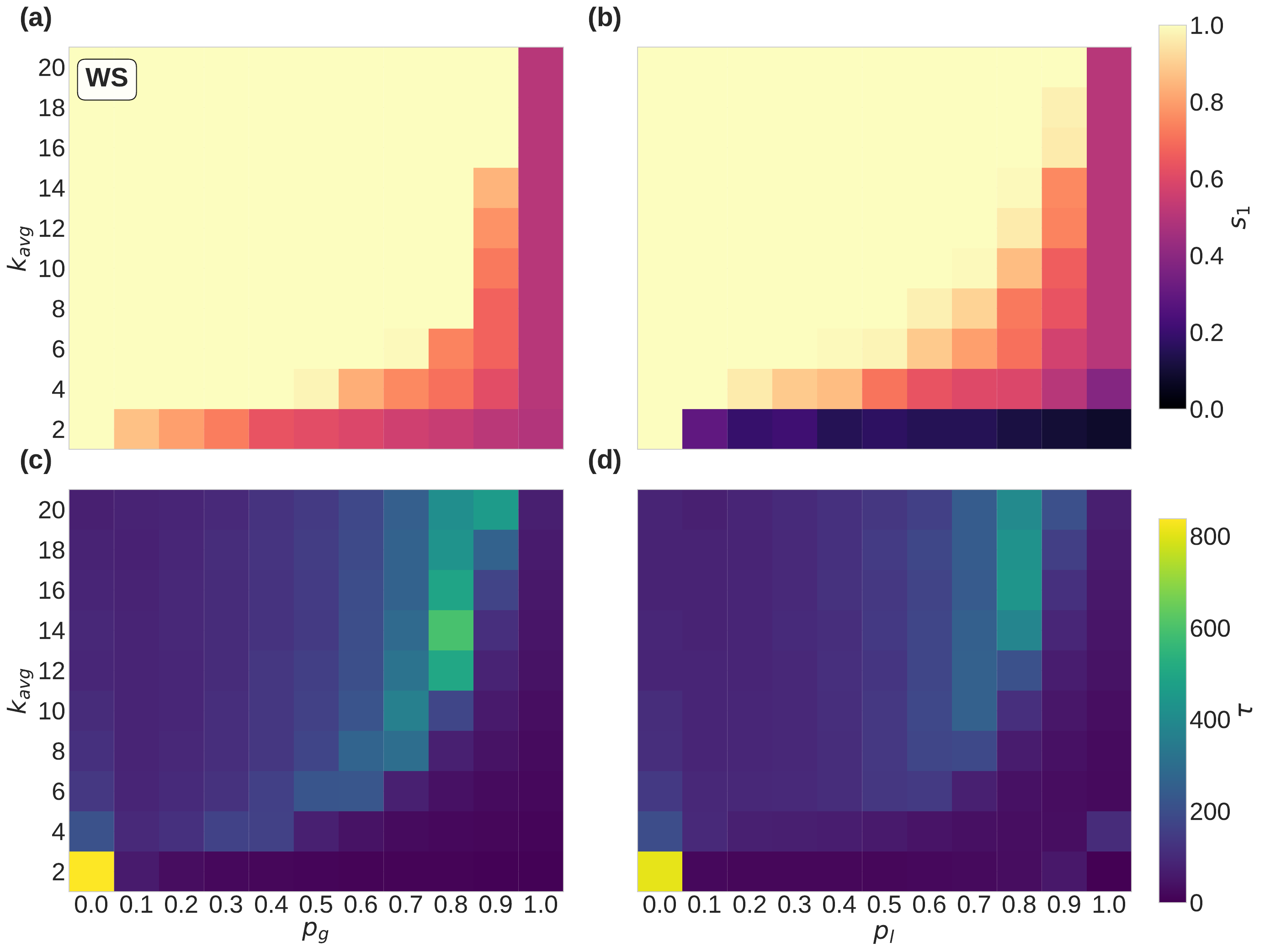}
		\caption{Heatmap of the biggest component size $s_1$ (top rows) and convergence time $\tau$ (bottom rows) for a WS with $N = 100$ and $k_{avg} = [2, 4,...,22]$. \textbf{(a), (c):} Global rewiring shows a clear transition point, increasing non-linearly with $k_{avg}$ for all topologies. \textbf{(b), (d):} Local rewiring leads to a smearing in that transition, with a smooth decrease of the biggest component and lower convergence times at the critical points of global rewiring}
		\label{fig:heatmap}
	\end{figure}

	\noindent Continuing with the phase transition behavior, figure (\ref{fig:heatmap}) provides an overview for convergence time per node $\tau$ and the final size of the biggest component $s_1$. For all three topologies, the local rewiring leads to a smoothing of the biggest component size due to isolated nodes or small ECs arising even with low rewiring probability $p_l$. The convergence time of the local rewiring is smaller around the phase transition region of the global rewiring with a slight shift of the maximal convergence time towards higher average degrees. Low degree in the WS network and pure imitation yields longer convergence times due to the near-stable clusters of different opinions in case of $k_{avg}$. Specifically, in case of the WS network with $k_{avg} = 2$ we have $C \approx 0$ for the overall clustering coefficient and the slow convergence is driven by the high distance between nodes \cite{albert2002statistical}.
	
    \noindent Like all models, ours involves various simplifications. It abstracts away psychological nuances of opinion formation and assumes homogeneous agent behavior, which may not capture real-world heterogeneity in interactions \cite{lee2014social, hall2018social}. Moreover, the model focuses on structural network effects and does not incorporate content-based features or temporal variations in user engagement \cite{alizadeh2020content}. Despite these limitations, the strength of our approach lies in its ability to isolate the impact of rewiring on social dynamics in a controlled environment. This allows for clear causal inferences that are often challenging to derive from empirical data alone \cite{vanderweele2013social} and enables a systematic test of different network topologies.
	
	\newpage
	
	\section{Conclusion}\label{Conclusions}
	
	Our computational study demonstrates that the design of social media recommendation algorithms, specifically the mechanism of link recommendation algorithms, plays a critical role in shaping opinion polarization and echo chambers. By comparing global (random) and local (friend-of-friend) homophilic rewiring within an adaptive voter model, we have isolated the impact of algorithmic design on network dynamics.
	\newline\newline
	\noindent Our key finding is that local, neighbor-based rewiring, mimicking common “people you may know” algorithms, significantly amplifies the formation of echo chambers compared to a global rewiring baseline. Local rewiring leads to higher levels of magnetization, corresponding to higher levels of polarization in combination with our homophily measure. This outcome is due to the lower rewiring probability required for fragmentation, which shatters the network into a much larger number of small, disconnected components. The effect is most pronounced in clustered networks with high levels of homophily in the rewiring dynamic. These results indicate that pre-existing social clustering and homophily can act as a precursor for algorithmic fragmentation.
	\newline\newline
	\noindent Furthermore, we observed that our algorithmic local rewiring has little effect on the critical behavior of the original AVM. Instead, it leads to a smoother change in the size of the largest network component due to the appearance of isolated nodes. Convergence time is affected only minimally, slowing down equilibration without shifting the point of convergence with respect to average degree or rewiring rate $p_{g,l}$. In addition, local rewiring $p_l$ shifts the degree distribution, creating a more unequal network with a higher prevalence of highly connected nodes across all standard network topologies.
	\newline\newline
	\noindent These results provide a mechanistic explanation for empirical concerns about social media algorithms fostering echo chambers \cite{kitchens2020understanding}. In particular, our model suggests that even at low rewiring probabilities, such algorithms can initiate fragmentation processes that lead to a divided social landscape.
    \newline\newline
    \noindent While the present study is not calibrated to empirical data, future work could use empirical network structures and interaction patterns to inform model parameters and assess whether similar fragmentation mechanisms arise in observed social networks. In this spirit, future work could address follower-based networks like X/Twitter by employing directed networks \cite{zschaler2012early}, include zealots or confident voters to test robustness against opinion inflexibility \cite{klamser2017zealotry, volovik2012dynamics}, and explore dynamic network sizes to reflect real-world platform growth and decline \cite{waller2021quantifying, bhat2019fixation}. Such extensions would deepen insights into polarization on evolving social platforms. In addition, the results could be enriched by complementary equation-based approaches to clearly identify critical behavior for local rewiring.
    
    \newpage
	
	\section{Appendix}\label{appendix}
    \subsection{Non-convergence of pure rewiring}\label{appendix:A}
	
	Table \ref{tab:convergence} provides a brief overview of convergence behavior for the special case mention in section \ref{Methodology}. For sparse graphs with a heterogeneous initial opinion distribution, subgraphs with different opinions can split off and due to no opinion imitation at $p_l = 1$, heterogeneous disconnected components can remain. 
		
	\begin{table}[h!]
		\centering
		\begin{tabularx}{0.95\textwidth}{|X|X|X|X|X|}
			\hline
			$k_{avg}$ & 2    & 4    & 6    & 8+    \\ \hline
			ER        & 0.91 & 0.97 & 1.00 & 1.00  \\
			WS        & 0.91 & 0.97 & 1.00 & 1.00  \\
			BA        & 0.92 & 0.98 & 1.00 & 1.00 \\ \hline
		\end{tabularx}
		\caption{Fraction of convergences in case of pure neighbour rewiring ($p = 1$) for different topologies. Convergence fails for lower average degrees $k_{avg} \le 4$ in less than \% of cases for ER and BA, but a high fraction of heterogeneous graphs remain for the WS network}
		\label{tab:convergence}
	\end{table}
	
	\vspace{-4.5em}
	
	\begin{figure}[H]
		\centering
		\includegraphics[width=0.45\linewidth]{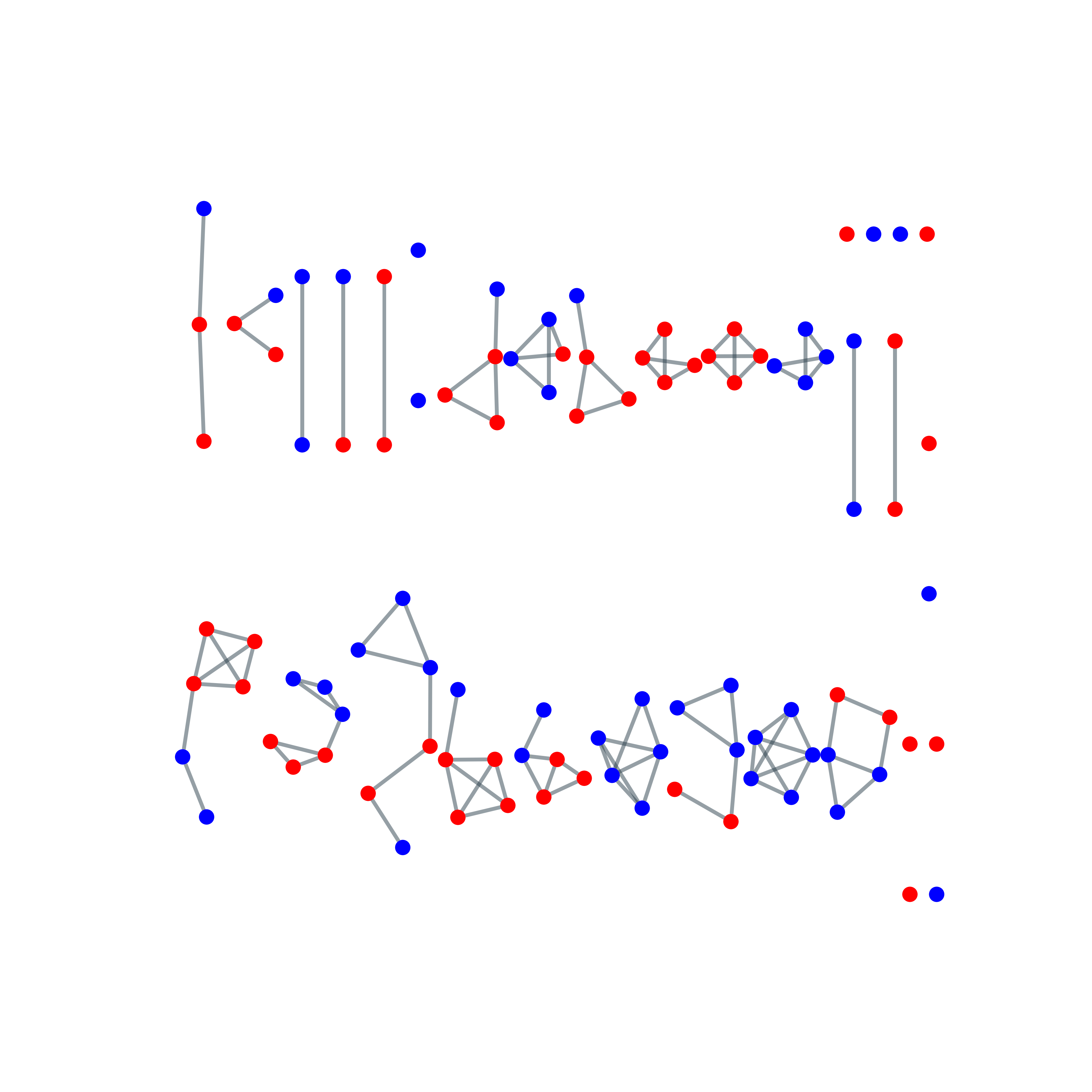}
		\includegraphics[width=0.45\linewidth]{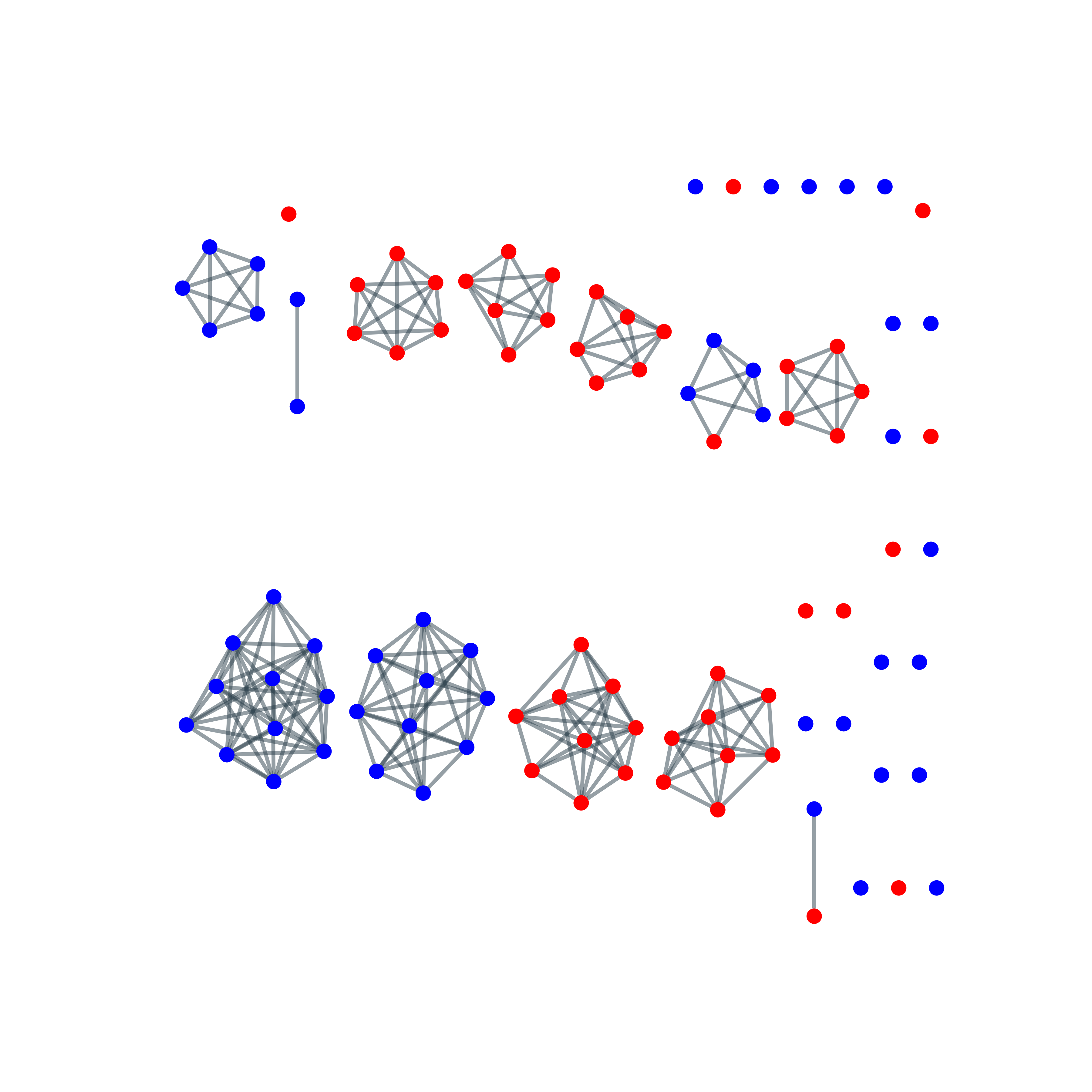}
		\caption{Final WS graphs with $N = 100$, $k_{avg} = [2,4]$ (left, right) with pure neighbour rewiring}
		\label{fig:example_graphs}
	\end{figure}
	
	\noindent This behavior can also be seen for pure random rewiring if the average degree reaches $k_{avg} \ge N/2$ as each node will be connected to more nodes than there are nodes with the same opinion. As we focus on social networks with a low average degree relative to the network size, this example was not considered in length.
	
	\subsection{Heatmaps for ER and BA}\label{appendix:B}
	
	\begin{figure}[H]
		\centering
		\includegraphics[width=0.7\textwidth]{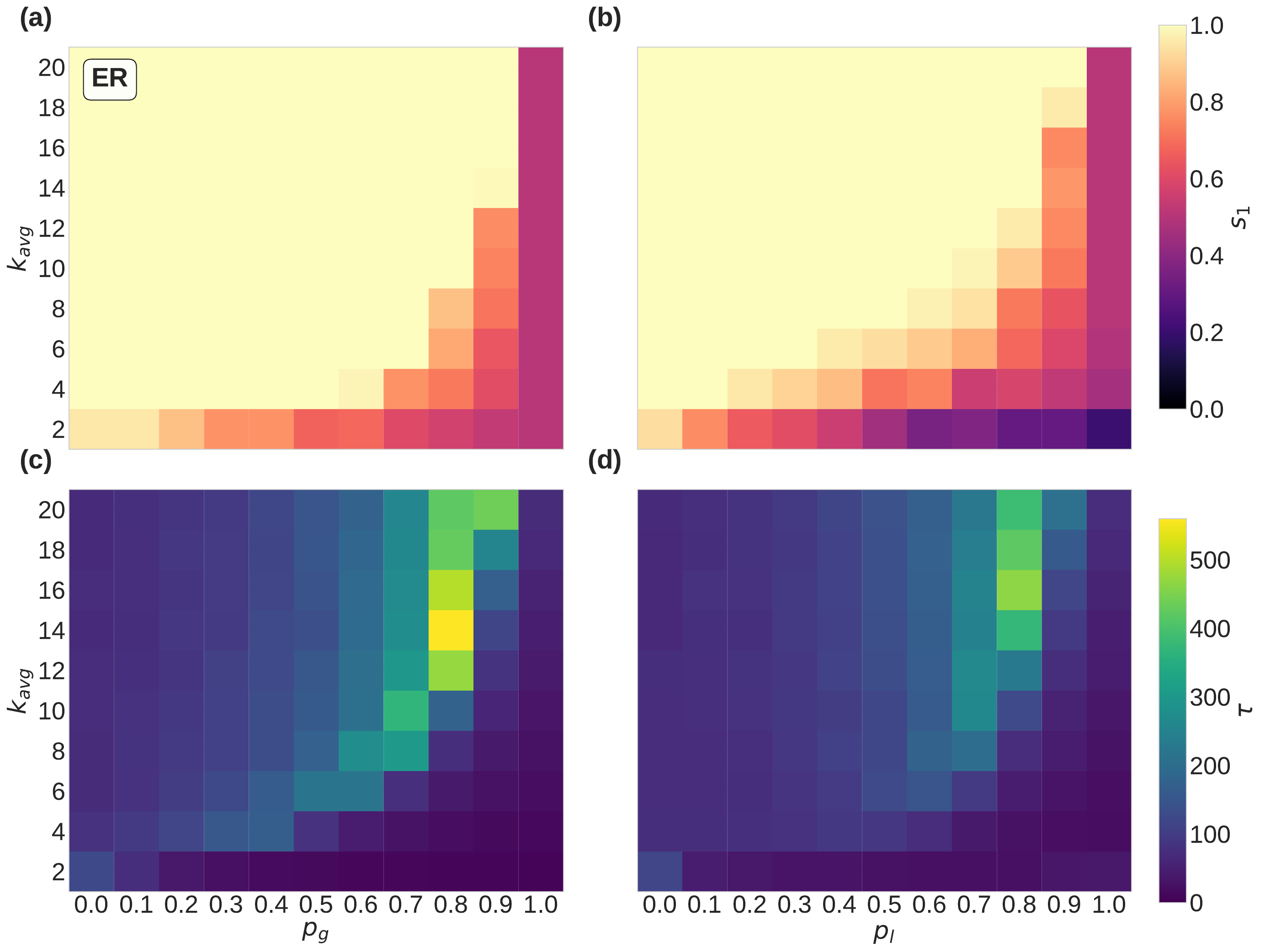}
		\includegraphics[width=0.7\textwidth]{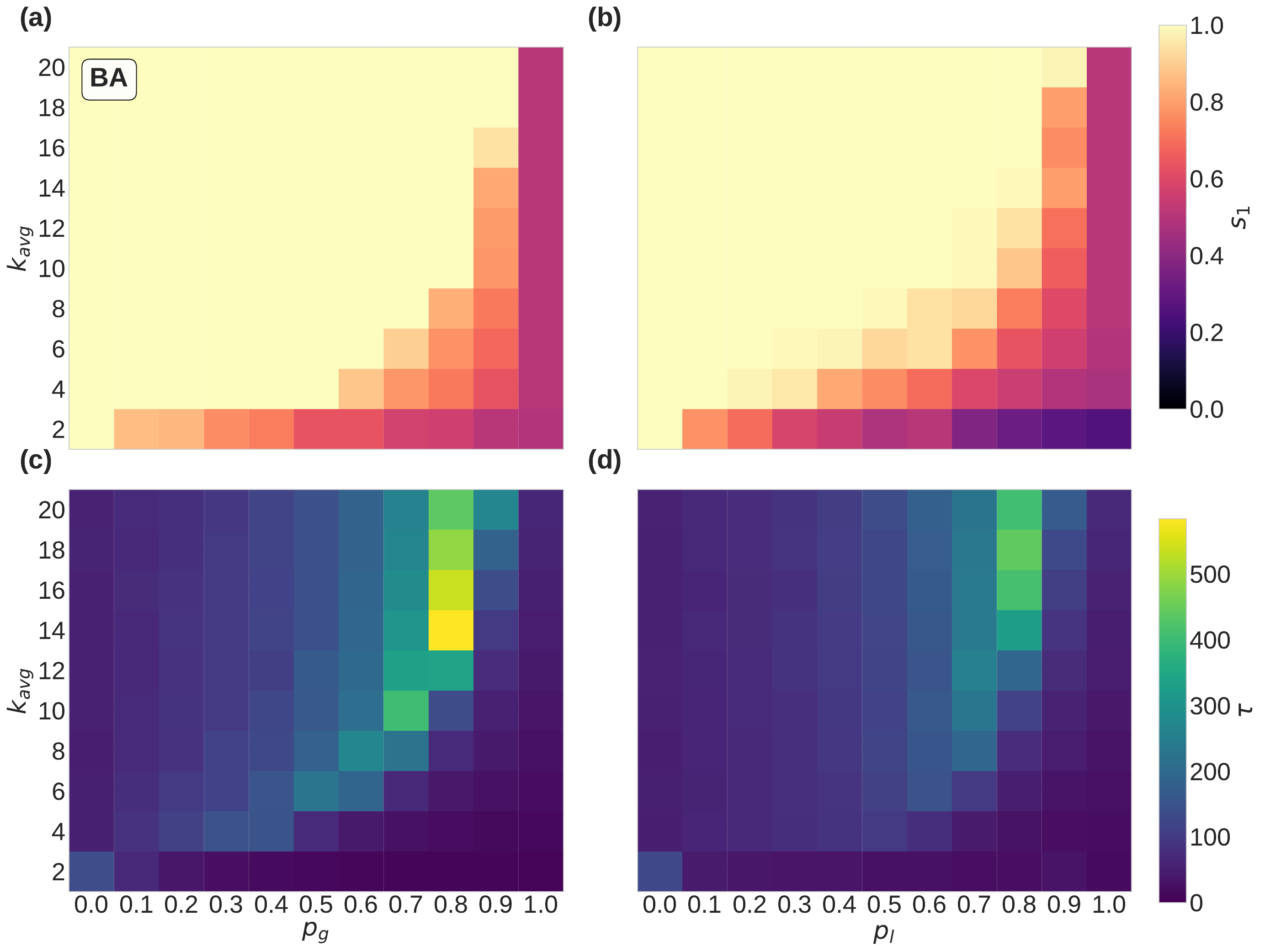}
		\caption{Heatmap of biggest component size $s_1$ (above) and convergence time $\tau$ (below) for an ER (top half) and BA (bottom half) network with $N = 100$ and $k_{avg} = [2, 4,...,22]$. \textbf{(a), (c):} Homophilic global rewiring shows a clear transition point, increasing non-linearly with $k_{avg}$ for all topologies. \textbf{(b), (d):} Homophilic local rewiring leads again here to a smearing in that transition, with a smooth decrease of the biggest component and lower convergence times at the critical points of global rewiring}
		\label{fig:heatmap2}
	\end{figure}
	
	\subsection{Homophily, magnetization and echo chambers for different degrees}\label{appendix:C}
	
	\begin{figure}[H]
		\centering
        \includegraphics[width=0.70\linewidth]{homophily_polarization_echo_chambers_WS_n100_k4_homo1.0.pdf}	
		\includegraphics[width=0.70\linewidth]{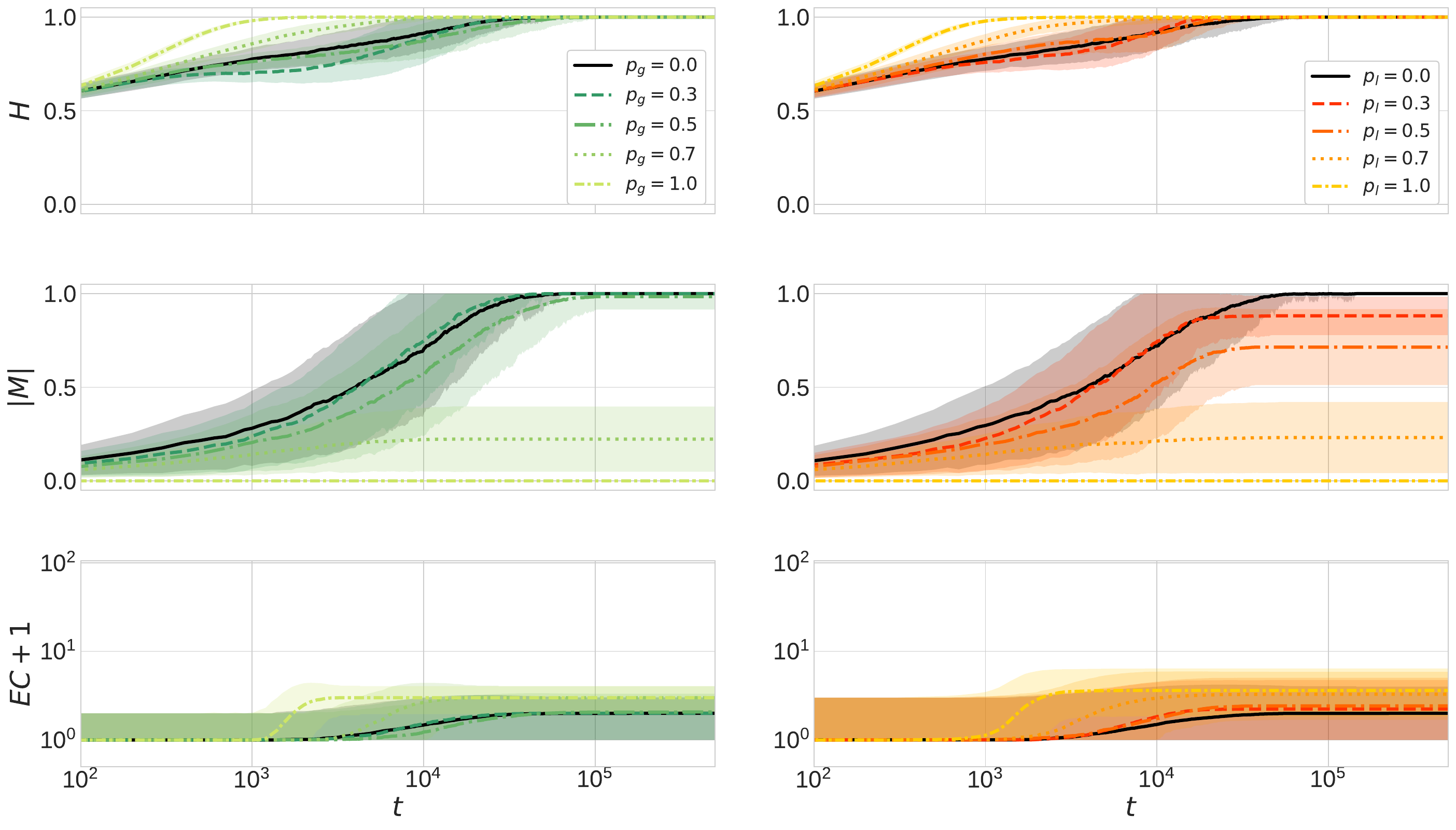}	\includegraphics[width=0.70\linewidth]{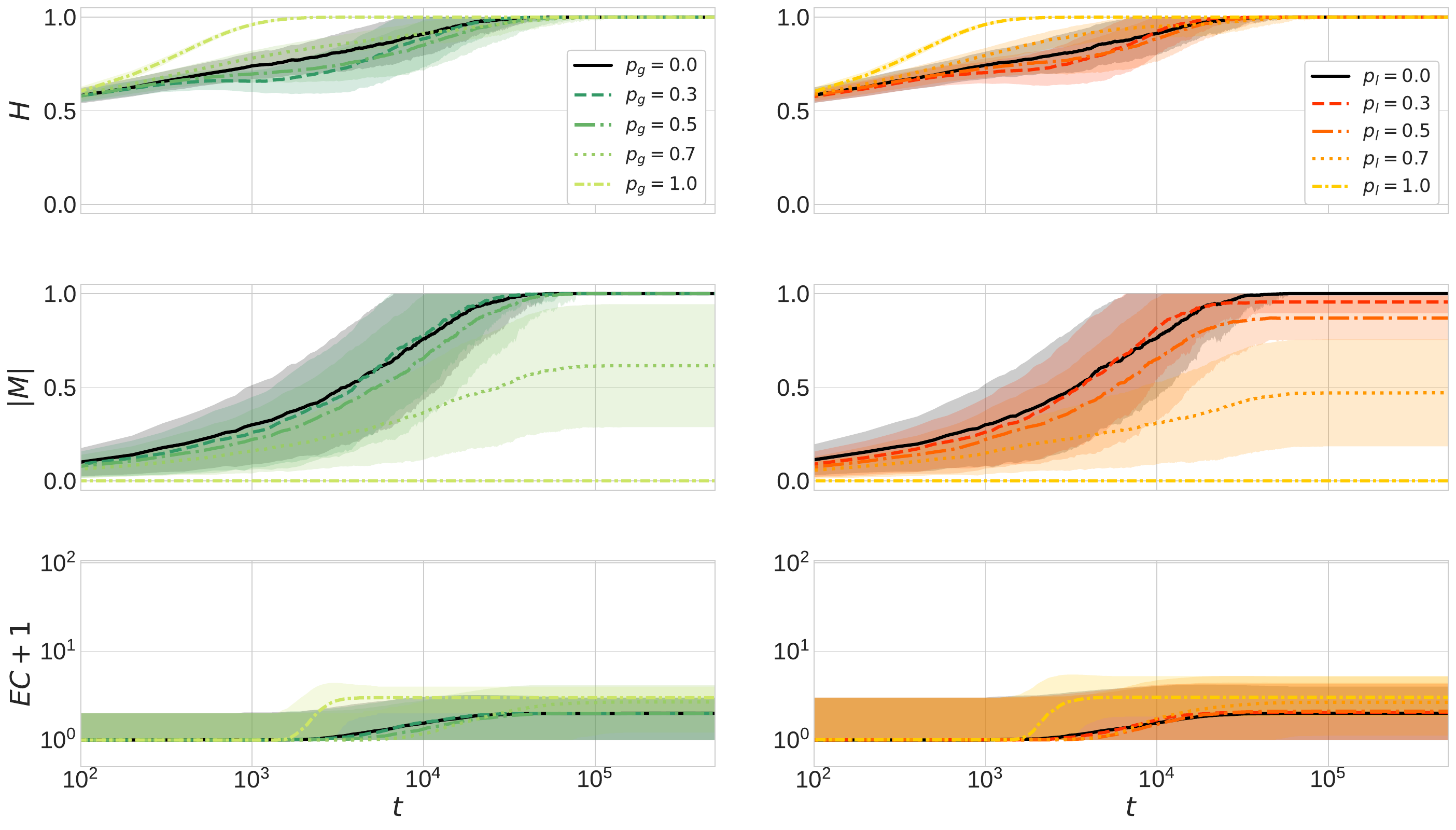}
		\caption{Overall homophily $H$, the absolute value of the magnetization $|M|$ and the shifted number of echo chambers $EC+1$ over time $t$ for a WS network with $N = 100$ and average degree $k_{avg} = 4$ (top), $k_{avg} = 6$ (top) and $k_{avg} = 8$ (bottom).}
		\label{fig:polarization2}
	\end{figure}
	
    \subsection{Homophily, magnetization and echo chambers for different levels of homophily}\label{appendix:D}

    \begin{figure}[H]
        \centering
        \includegraphics[width=0.65\linewidth]{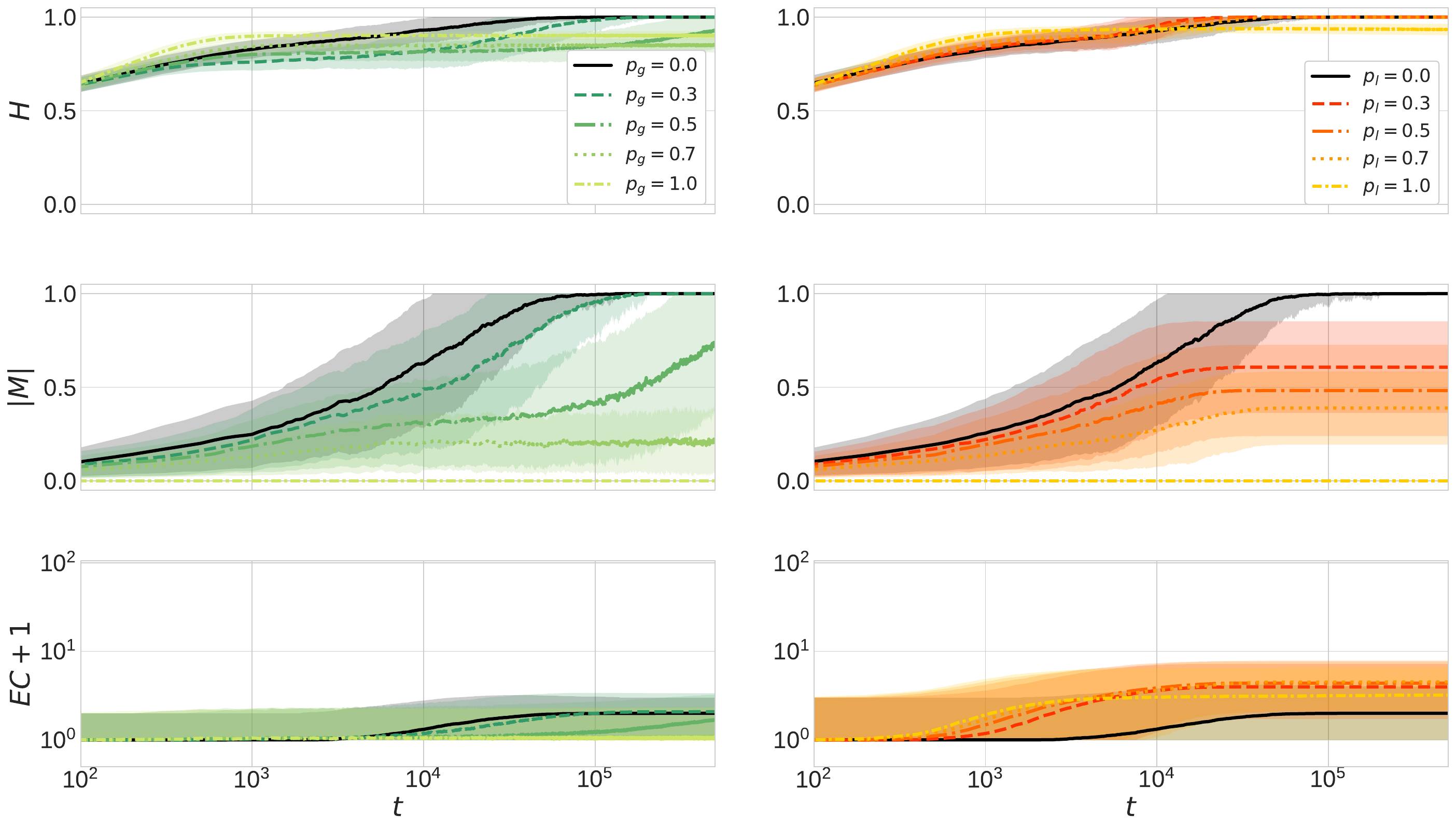}
        \includegraphics[width=0.65\linewidth]{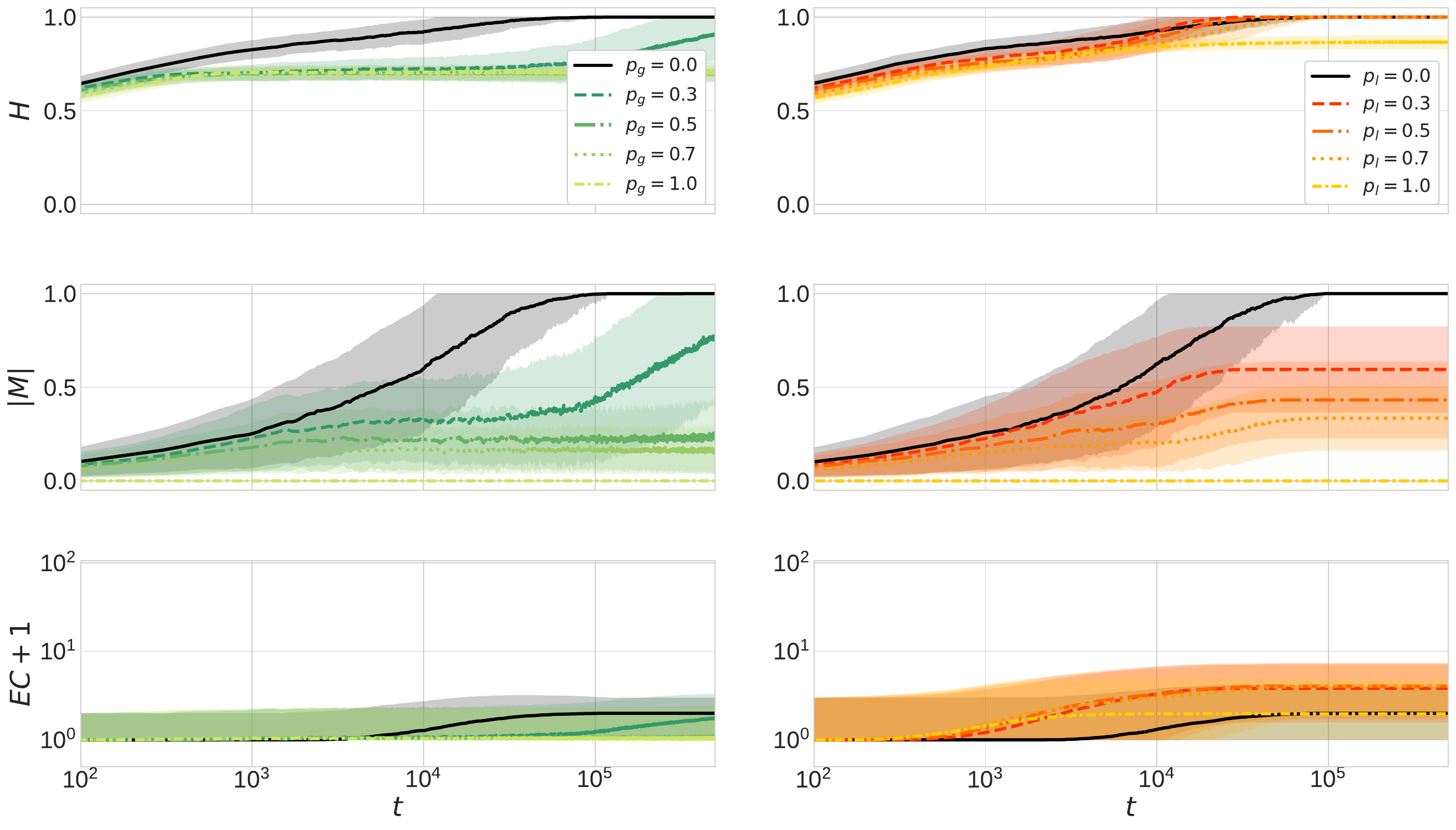}
        \includegraphics[width=0.65\linewidth]{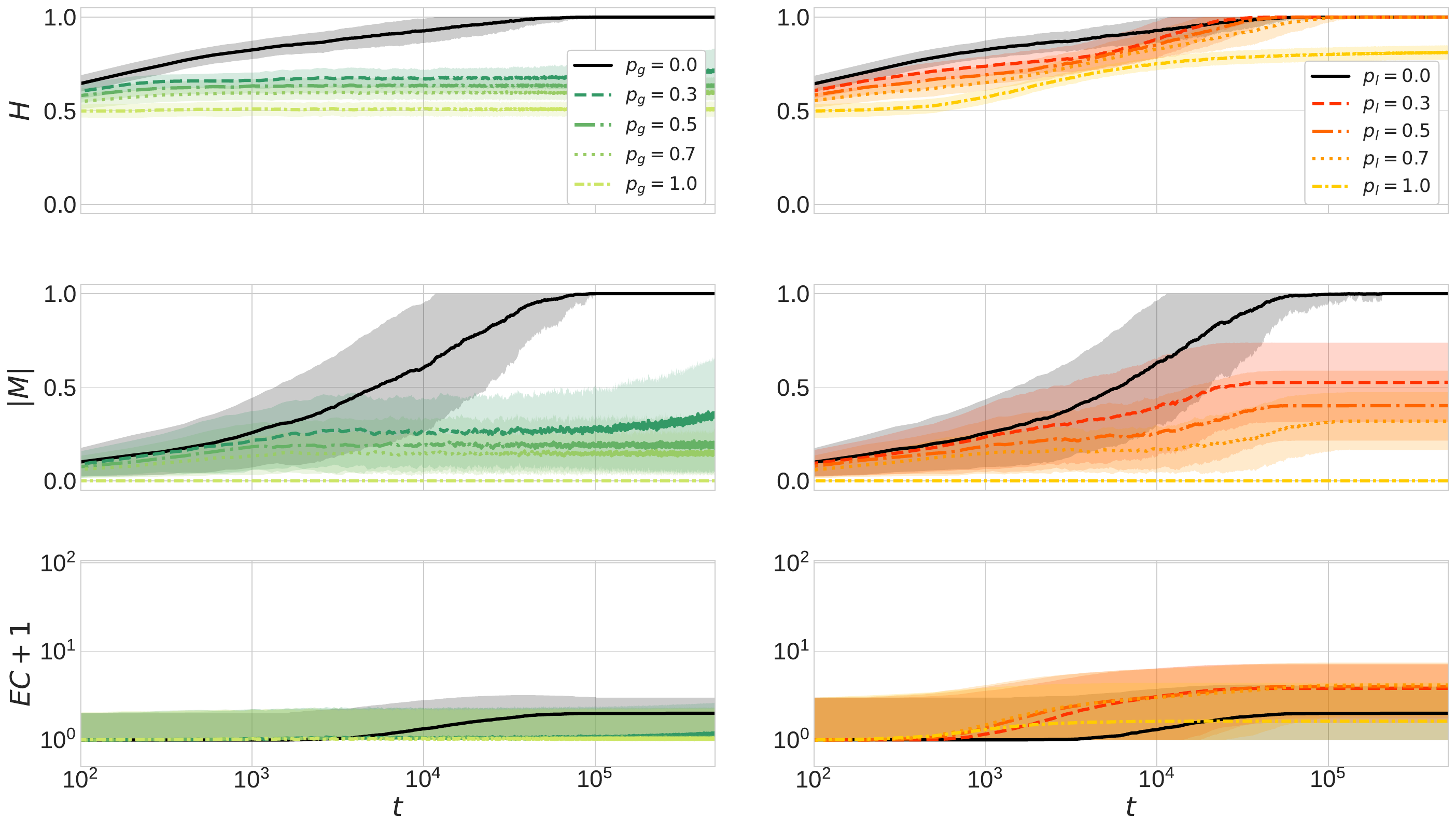}
        \caption{Overall homophily $H$, absolute value of magnetization $|M|$ and the shifted number of echo chambers $EC+1$ over time $t$ for a WS network with $N = 100$, average degree $k_{avg} = 4$ and homophily $p_h = 0.9$ (top), $p_h = 0.7$ (middle) and $p_h = 0.5$ (bottom).}
        \label{fig:polarization3}
    \end{figure}
    Figure (\ref{fig:polarization3}) shows that the number of echo chambers $EC$ declines for $p_l = 1$ as soon as homophily $p_h$ is reduced, but stays roughly constant for $0<p_l <1$ across homophily values $p_h$.

    \noindent \textbf{Data Availability} No data associated in the manuscript.
    
    \section*{Declarations}
    \textbf{Conflict of interest} On behalf of all authors, the corresponding author states that there is no conflict of interest. 
    
    \bibliographystyle{plain}
    \bibliography{bibliography}
    	
\end{document}